\documentclass[fleqn,usenatbib]{mnras}
\usepackage{newtxtext,newtxmath}
\usepackage[T1]{fontenc}

\DeclareRobustCommand{\VAN}[3]{#2}
\let\VANthebibliography\thebibliography
\def\thebibliography{\DeclareRobustCommand{\VAN}[3]{##3}\VANthebibliography}

\usepackage{graphicx}	% Including figure files
\usepackage{amsmath}	% Advanced maths commands
\usepackage{url}
\newcommand{\angs}{\mathring{A}} % angstrom symbol for overlead
\title[Polarimetry of young ZTF supernovae]{RINGO3 polarimetry of very young ZTF supernovae}
\author[Maund et al.]{
J. R. Maund$^{1}$\thanks{E-mail: j.maund@sheffield.ac.uk}, Y. Yang$^2$, I. A. Steele$^{3}$, D. Baade$^4$, H. Jermak$^{3}$, S. Schulze$^2$,\newauthor 
R. Bruch$^{2}$, A. Gal-Yam$^{2}$, P. A. H\"oflich$^{5}$, E. Ofek$^{2}$, X. Wang$^{6, 7}$, M. Amenouche$^{8}$, \newauthor
R. Dekany$^{9}$, F. J. Masci$^{9}$, R. Riddle$^{9}$ \& M.~T. Soumagnac$^{2, 10}$
\\
$^{1}$Department of Physics and Astronomy, University of Sheffield, Hicks Building, Hounsfield Road, Sheffield, S3 7RH, UK \\
$^{2}$ Department of Particle Physics and Astrophysics, Weizmann Institute of Science, 234 Herzl St., 71600 Rehovot, Israel \\
$^{3}$ Astrophysics Research Institute, Liverpool John Moores Univerisity, 146 Brownlow Hill, Liverpool, L3 5RF, UK \\
$^4$ European Organisation for Astronomical Research in the Southern Hemisphere (ESO), Karl-Schwarzschild-Str. 2, 85748 Garching b.\ M\"unchen, Germany\\
$^{5}$ Department of Physics, Florida State University, 77 Chieftan Way, Tallahassee, FL 32306, USA\\
$^{6}$ Physics Department, Tsinghua University, Beijing, 100084, China\\
$^{7}$ Beijing Planetarium, Beijing Academy of Science and Technology, Beijing, 100089, China\\
$^{8}$ Universit{\'e} Clermont Auvergne, CNRS/IN2P3, LPC, F-63000 Clermont-Ferrand, France\\
$^{9}$ California Institute of Technology, 1200 E. California Blvd, Pasadena, CA 91125, USA\\
$^{10}$Lawrence Berkeley National Laboratory, 1 Cyclotron Road, Berkeley, CA 94720, USA\\
}

\date{Accepted XXX. Received YYY; in original form ZZZ}

\pubyear{2021}

\begin{document}
\label{firstpage}
\pagerange{\pageref{firstpage}--\pageref{lastpage}}
\maketitle

% Abstract of the paper
\begin{abstract}
The early phases of the observed evolution of the supernovae (SNe) are expected to be dominated by the shock breakout and ``flash" ionization of the surrounding circumstellar medium.  This material arises from the last stages of the evolution of the progenitor, such that photometry and spectroscopy of SNe at early times can place vital constraints on the latest and fastest evolutionary phases leading up to stellar death.  These signatures are erased by the expansion of the ejecta within $\sim 5$ days after explosion.
Here we present the earliest constraints, to date, on the polarization of ten transients discovered by the Zwicky Transient Facility (ZTF), between June 2018 and August 2019.  Rapid polarimetric followup was conducted using the Liverpool Telescope RINGO3 instrument, including 3 SNe observed within $<1$ day of detection by the ZTF.  The limits on the polarization within the first 5 days of explosion, for all SN types, is generally $<2\%$, implying early asymmetries are limited to axial ratios $>0.65$ (assuming an oblate spheroidal configuration).  We also present polarimetric observations of the Type I Superluminous SN 2018bsz and Type II SN 2018hna, observed around and after maximum light.
\end{abstract}

% Select between one and six entries from the list of approved keywords.
% Don't make up new ones.

%%%%%%%%%%%%%%%%%%%%%%%%%%%%%%%%%%%%%%%%%%%%%%%%%%

%%%%%%%%%%%%%%%%% BODY OF PAPER %%%%%%%%%%%%%%%%%%

%%%%%%%%%%%%%%%%%%%%%%%%%%%%%%%%%%%%%%%%%%%%%%%
%OBSERVATIONS
%OBSERVATIONS
%INTRODUCTION
%%%%%%%%%%%%%%%%%%%%%%%%%%%%%%%%%%%%%%%%%%%%%%%

\section{Introduction}
\label{sec:intro}
With the advent of deep, wide-field and high-cadence surveys, it has been possible to discover new supernovae (SNe) within hours of explosion.  Rapid followup observations of these SNe has provided a new insight into the stellar origins of these explosion.  Early photometric observations of the emergence of the explosion shock has, in a number of cases \citep[e.g.][]{2008natur.453..469s,2010apj...724.1396o, 2016apj...820...23g, 2017apj...848....8r, 2018natur.554..497b}, revealed a behaviour that cannot be explained by just the shock breaking out of the stellar surface \citep{2017hsn..book..967w}.  Instead, the early, rapid rise in brightness can be greatly affected by the presence of circumstellar material (CSM).  

For a subset of Type Ia SNe, early photometric observations have revealed an ultraviolet excess, which has been interpreted as the shock interaction between the ejecta and a companion star \citep{2015natur.521..328c, 2017apj...845l..11h}.  Early spectroscopy has revealed the presence of ``flash ionized" species, such as He {\sc ii} and N {\sc iv}, which correspond to the very quick ionization of the surrounding CSM \citep{2016apj...818....3k,2019mnras.483.3762k}.  In the case of SN~2013cu, early observations by \citet{2014natur.509..471g} were able to reveal the presence of a Wolf-Rayet-like wind immediately around the progenitor \citep{2007ara&a..45..177c}.  Using flash spectroscopy, \citet{2017natph..13..510y} showed that some SN progenitors may exhibit enhanced levels of mass loss as pre-supernova instabilities become important in the final years before explosion.  \citet{2020arxiv200809986b} report that at least $30\%$ of hydrogen-rich SNe exhibit such features in very early observations, consistent with elevated mass-loss just prior to explosion.  Despite the power of early time observations to place important constraints on the nature of the progenitor system, the observable signatures disappear by $\sim 5$ days, after the ejecta overrun the immediately surrounding dense CSM \citep{2014natur.509..471g}.  Early-time observations are, therefore, very sensitive to the very last phases of stellar evolution.

The early-time optical observations of very young supernovae have, predominantly, been in the form of photometry and spectroscopy.  Polarimetry has been established as a sensitive probe of the presence of departures from spherical symmetry in SN explosions \citep{2008ara&a..46..433w}.  Observations around maximum light and at later epochs, as the photosphere recedes into the ejecta with time, have revealed important clues to the physics of the explosions responsible through their imprint on the geometry of the ejecta. In general, core-collapse (CC) SNe show increasing degrees of polarization with time indicating the ejecta becoming more asymmetric the closer to the origin of the explosion.  On the other hand, Type Ia SNe exhibit the opposite behaviour appearing to become progressively more spherical closer to origin of the explosion \citep{2008ara&a..46..433w}.

The application of polarimetry to SNe at very early times, however, has been limited, with the earliest spectropolarimetric observation of a Type Ia SN published to date, occurring at only $\sim 5$ days after explosion \citep{2020apj...902...46y}.  Given the power of polarimetry to probe the 3D structures of these events, at early times it has the potential to probe the shape of the progenitor system, including the nature of the mass loss prior to explosion.   Indeed, as shown by \citet{2014mnras.442.1166m} and \citet{2017mnras.470.1491r}, in the context of SN 2009ip, polarimetry can provide constraints on both the 3D physical characteristics of the explosions and their interaction with the CSM, that are not accessible with ordinary photometric and spectroscopic observations.

A major difficulty with conducting rapid polarimetric followup of young supernovae is establishing the connection between the discovery surveys and facilities with an appropriate polarimetric observing capability.  Here, we report a pathfinder campaign, using the 2.0m Liverpool Telescope \citep{2004spie.5489..679s} and the RINGO3 polarimeter \citep{2012spie.8446e..2ja}, to acquire early-time observations of explosive transients; in particular those discovered by the Zwicky Transient Facility \citep[ZTF;][]{2019pasp..131a8002b,  2019pasp..131g8001g, 2019aas...23313106g}.  The RINGO3 instrument and its predecessors \citep{2006spie.6269e..5ms, 2010spie.7735e..49s} were designed for the rapid followup of the Gamma Ray Burst afterglows \citep{2007sci...315.1822m,2009natur.462..767s,2013natur.504..119m}, exploiting the flexibility to rapidly observe new targets afforded by the robotic nature of the Liverpool Telescope.  Given the location of ZTF at Palomar Observatory, California, USA, it is possible to trigger polarimetric followup observations with the Liverpool Telescope (La Palma, Canary Islands, Spain) within $<24$ hours of discovery.

%%%%%%%%%%%%%%%%%%%%%%%%%%%%%%%%%%%%%%%%%%%%%%%
%OBSERVATIONS
%OBSERVATIONS
%OBSERVATIONS
%%%%%%%%%%%%%%%%%%%%%%%%%%%%%%%%%%%%%%%%%%%%%%%

\section{Observations}
\label{sec:obs}
\subsection{Data Acquisition \& Reductions}
\label{sec:dataacq}
All observations of the target SNe were conducted with the RINGO3 instrument\footnote{\url{https://telescope.livjm.ac.uk/TelInst/Inst/RINGO3/}} mounted on the Liverpool Telescope.  RINGO3 operates with three separate channels, each with its own camera, with the light split by wavelength using two dichroics.  The three channels are: ``d'' covering $7700 - 10\,000\mathrm{\angs}$,``e'' covering  $3500 - 6400\mathrm{\angs}$ and ``f'' covering $6500 - 7600\mathrm{\angs}$; following the nomenclature of \citet{jermakphd}, we will refer to these band passes as $r^{\ast}$, $b^{\ast}$ and $g^{\ast}$, respectively.  Each camera has a slightly different plate scale: $0.43\arcsec$, $0.44\arcsec$ and $0.49\arcsec$ for the $b^{\ast}$, $g^{\ast}$ and $r^{\ast}$ bands, respectively.  Each RINGO3 channel has its own $512 \times 512\,\mathrm{px}$ Electron Multiplying CCD which have, for the type of observations considered here, negligible noise associated with readout.

RINGO3 uses a rotating polaroid ($\sim 0.4$Hz) to sample all of the components of the Stokes parameters at eight separate polaroid position angles.  Each camera produces, therefore, 8 exposures in 2.3 seconds (i.e. 24 exposures total for the 3 cameras). The RINGO Data Reduction Pipeline\citep{arnoldphd}\footnote{\url{https://telescope.livjm.ac.uk/TelInst/Inst/RINGO3/\#pipeline}} creates a series of mean stacked images: one for the entire duration of the observation and a series of mean stacked images for each minute of the observation.  For this study we only consider the mean stacked image constructed from all exposures acquired at a given epoch.

The reduced images of the science targets and the zero-polarization and highly-polarized standard calibration stars were retrieved from the Liverpool Telescope Archive\footnote{\url{https://telescope.livjm.ac.uk/cgi-bin/lt\_search}}.  

\subsection{Data Analysis Workflow}
\label{sec:dawf}
In order to analyse the data, we created a bespoke package to process all the observations and, ultimately, derive the linear Stokes parameters for the science targets.  The approach to the analysis follows those presented by \citet{jermakphd} and \citet{2019mnras.482.4057m}.

The data were first sorted into discrete datasets, containing all 24 files corresponding to one individual observation.  For each dataset and for each camera, source detection was conducted on the image at the first rotor position.  These positions were then used to conduct photometry on the images at all 8 rotor positions.   Due to extreme vignetting for all three cameras, source detection was not conducted in the four corners of the images (corresponding to areas of $128 \times 128\mathrm{px}$).   Aperture photometry was conducted using the {\it python} {\tt photutils}\footnote{\url{https://pypi.org/project/photutils/}} package.  For the bright standards, we used a fixed aperture of radius 8 pixels.  For the science targets, aperture sizes were selected to match the full width at half maximum to balance possible contamination from nearby stars or enhanced background (e.g. host galaxy), but maximise the signal-to-noise.  In the event that it was not possible for the Liverpool Telescope Data Reduction Pipeline to confidently establish the World Coordinate System for each image, the target (or targets) of interest in each image were selected by hand.  For all targets, the intensity ($I$) and the normalized linear Stokes parameters ($q$ and $u$) were calculated from the measured fluxes $f_{i}$ at each rotor position $i$. Following the prescription of \citet{2002A&A...383..360C}, the intensities corresponding to each of the three Stokes parameters is given by:

\begin{equation}
\begin{aligned}
S_{I} & = \sum^{8}_{i = 1}f_{i}\\
S_{q} & = f_{2} + f_{3} + f_{6} + f_{7}\\
S_{u} & = f_{1} + f_{2} + f_{5} + f_{6} \\
\end{aligned}
\label{eq:rawstokes}
\end{equation}

for which the normalised, instrumental Stokes parameters are:

\begin{equation}
\begin{aligned}
q_{inst} & = \pi \left(\frac{1}{2} - \frac{S_{q}}{S_{I}} \right)\\
u_{inst} & = \pi \left(\frac{S_{u}}{S_{I}} - \frac{1}{2} \right)
\end{aligned}
\label{eq:measstokes}
\end{equation}

In order to correctly propagate the photometric uncertainties, we used Monte-Carlo sampling to create N = 10000 samples from the distribution $N(f_{i}, (\Delta f_{i})^2)$ and carried these distributions through Equations \ref{eq:rawstokes} and \ref{eq:measstokes}, to derive the corresponding distributions for $q_{inst}$ and $u_{inst}$.  
Observations of zero and highly polarized standards were used to remove any instrumental polarization signature.  The standard stars observed as part of the Liverpool Telescope RINGO3 standard calibration plan are derived from \citet{1992aj....104.1563s}.  The instrumental polarization offset $(q_{0}, u_{0})$ was calculated using the observations of the zero-polarization standard stars, such that:
\begin{equation}
\begin{aligned}
q^{\prime} & = 1.14 (q_{inst} - q_{0})  \\
u^{\prime} & =  u_{inst} - u_{0}
\end{aligned}
\end{equation}
where the factor 1.14 corrects for elliptical distortion of the polarization circle of a constant polarization source in the $qu$ plane \citep{arnoldphd}.

The polarization properties of the highly polarized standards in the appropriate RINGO3 wavelength channels are given in Table \ref{tab:app:hpol}, as previously used by \citet{2019mnras.482.4057m}.  We used reported polarization values for 7 standards from the ultraviolet to the infrared \citep[in the $UBVRIJHK$ passbands][]{1992aj....104.1563s}, employing a fourth order polynomial to calculate the brightness-weighted polarization over the wavelength ranges corresponding to the three RINGO3 channels.

\begin{table}
    \centering
    \caption{Polarized standards from \citet{1992aj....104.1563s} in the RINGO3 channels.\label{tab:app:hpol}}
    \begin{tabular}{ccccccccc}
    \hline
    Standard & \multicolumn{2}{c}{$b^{\ast}$/``e"}& & \multicolumn{2}{c}{$g^{\ast}$/``f"}& &\multicolumn{2}{c}{$r^{\ast}$/``d"}\\
    \cline{2-3}\cline{5-6}\cline{8-9}
     & $p(\%)$ & $\theta$ ($^{\circ}$)& &$p(\%)$ & $\theta$ ($^{\circ}$)& &$p(\%)$ & $\theta$ ($^{\circ}$)\\
    \hline
    BD$\,+25\,727$ & 5.99 & 31.2 & &6.13 & 31.5 & & 5.22 & 31.7\\
    BD$\,+59\,389$ &6.40 & 98.2  & &6.27 & 98.2 & & 5.45 & 98.2\\
    BD$\,+64\,106$ &5.48 & 96.9  & &5.00 & 96.8 & & 4.89 & 96.7\\
    HD$\,$155528   &4.80 & 91.9  & &4.80 & 91.9 & & 4.80 & 91.8\\
    HD$\,$215806   &1.80 & 66.6  & &1.74 & 69.0 & & 1.40 & 70.8\\
    Hiltner 960    &5.61 & 55.2  & &4.98 & 54.3 & & 4.19 & 53.5\\
    VI Cyg 12      &8.42 & 115.6 & &8.42 & 115.6 & & 8.42 & 115.6 \\
    \hline
    \end{tabular}
\end{table}
For highly polarized stars, the observed polarization angle, in the instrument coordinate frame, is given as:
\begin{equation}
\theta_{obs} = \frac{1}{2} \arctan \left(\frac{u^{\prime}}{q^{\prime}} \right)
\end{equation}
from which the rotation offset of the instrument can be determined by:
\begin{equation}
K = \theta_{0} - (ROTSKYPA - \theta_{obs})
\end{equation}
where $\theta_{0}$ is the previously determined polarization angle for the standard star in the reference catalogue and $ROTSKYPA$ is the instrument rotation angle.  $K$ is, therefore, the relative offset of the polaroid positions, with respect to the standard astronomical definition of the Stokes parameter coordinate system ($+q$ aligned with North and a polarizaton angle of $0^{\circ}$), without any rotation of the instrument.  The total degree of observed polarization was calculated as:
\begin{equation}
p_{obs}  = \sqrt{q^{\prime\,2} +u^{\prime\,2}}
\end{equation}
Through comparison with previously catalogued values of the polarization for the highly-polarized standard stars ($p_{0}$), the degree of instrumental ``depolarization” was derived as:
\begin{equation}
D = \frac{p_{0}}{p_{obs}}
\end{equation}
We note that this definition of the instrumental depolarization is the inverse to that used by \citet{2016mnras.458..759s}, however the two approaches are otherwise equivalent.
For the science targets, $q^{\prime}$ and $u^{\prime}$ were used to calculate the intrinsic polarization angle:
\begin{equation}
\theta_{0} = K + ROTSKYPA - \frac{1}{2}\arctan\left(\frac{u^{\prime}}{q^{\prime}} \right)
\end{equation}
and the intrinsic degree of polarization:
\begin{equation}
p_{0} = D \times p_{obs}
\end{equation}

The degree of polarization was further corrected for bias using the Modified ASymptotic (MAS) estimator of true polarization $p_{MAS}$ \citep{2014mnras.439.4048p}.  We follow \citet{2019mnras.482.5023h}, by characterising polarization measurements with $p_{MAS}/\sigma_{p} < 3$ as non-detections, and quote the 95\% upper limit.

\subsection{The Stability of the RINGO3 Instrument}
\label{sec:ringostab}
To establish a baseline calibration for each science observation we utilised the zero- and high-polarization standards, observed as part of the standard RINGO3 calibration plan, from the night of and the nights before and after the science observation.  As a single-beam polarimeter \citep{2012spie.8446e..2ja}, in which orthogonal polarization components are measured in series, instrumental and background effects may be additive and not completely removed through the calculations presented in Equations \ref{eq:rawstokes} and \ref{eq:measstokes}.  The determination of the instrumental polarization calibration parameters $q_{0}$, $u_{0}$, $D$ and $K$ may also be limited by the level of the sky background, the seeing and the throughput of each individual channel \citep{2016mnras.458..759s}.  RINGO3 also uses two dichroics (to separate the three separate wavelength channels) and a depolarizing Lyot prism, for which it is estimated the minimum total systematic uncertainty is $\sim 0.5\%$ \citep{jermakphd}. 

\begin{figure}
    \centering
    \includegraphics[width=7cm]{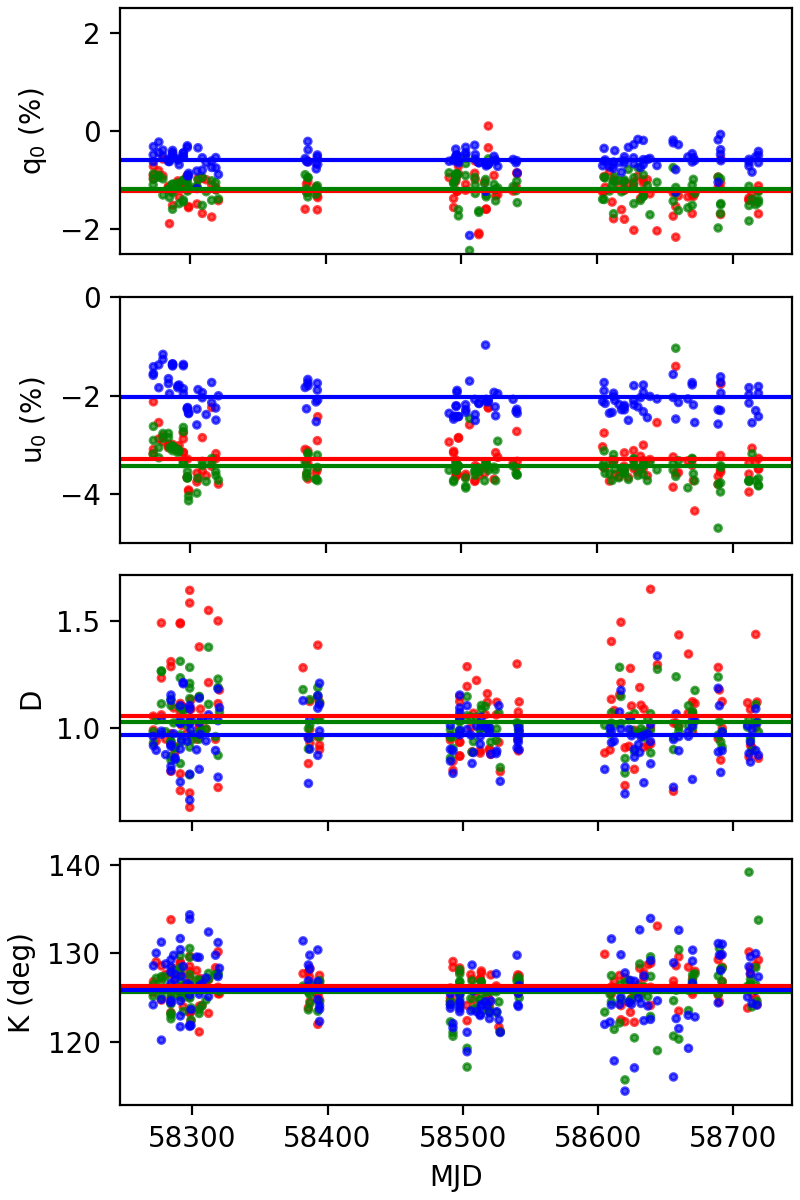}
    \caption{The instrumental polarization parameters derived from each zero- and high-polarization standard observed over the survey period.  The points are colour-coded according to wavelength channel (blue=$b^{\ast}$, green=$g^{\ast}$ and red=$r^{\ast}$).  Horizontal lines in each panel indicate the mean derived instrumental parameter (as summarised in Table \ref{tab:ringostab})}
    \label{fig:ringostab}
\end{figure}

The measured instrumental polarization parameters, for each polarization standard star observed, are shown in Figure \ref{fig:ringostab}.  The mean and standard deviation of the instrumental polarization parameters of RINGO3, for the survey period, are summarised in Table \ref{tab:ringostab}.  We also calculated the intra-night standard deviation of the instrumental polarization parameters $\sigma_{N}$ (and in Table \ref{tab:ringostab} we report the average over all nights).  In general, we find that the limiting systematic precision of RINGO3 is consistent with previous estimates \citep{jermakphd, 2016mnras.458..759s}.  Although the calibration of RINGO3 is relatively stable over the period of the survey, there is some structure present in Figure \ref{fig:ringostab} (e.g. around $MJD58300$, which coincided with the cleaning of a mirror in the optical path) which requires applying calibrations derived over short timescales (rather than an average derived over the entire length of the survey).

\begin{table*}
\caption{Average properties of the RINGO3 instrumental polarization between June 2018 and August 2019. \label{tab:ringostab}}
\begin{tabular}{cc|ccc|ccc|ccc|ccc}
\hline\hline
  \multicolumn{2}{c}{Channel} & $\overline{q_{0}}$ & $\sigma (q_{0})$ & $\overline{\sigma_{N} (q_{0})}$ &
$\overline{u_{0}}$ & $\sigma (u_{0})$ & $\overline{\sigma_{N} (u_{0})}$ &
$\overline{D}$ & $\sigma (D)$ & $\overline{\sigma_{N} (D)}$ &
$\overline{K}$ & $\sigma (K)$ & $\overline{\sigma_{N} (K)}$ \\

\hline

$b^{\ast}$ & e & -0.58 & 0.24 & 0.10 & -2.02 & 0.39 & 0.15 & 0.97 & 0.11 & 0.07 & 125.92 & 3.55 & 2.19 \\
$g^{\ast}$ & f & -1.18 & 0.28 & 0.14 & -3.44 & 0.38 & 0.11 & 1.03 & 0.10 & 0.05 & 125.62 & 2.70 & 1.16 \\
$r^{\ast}$ & d & -1.22 & 0.40 & 0.15 & -3.28 & 0.44 & 0.19 & 1.06 & 0.20 & 0.11 & 126.38 & 2.14 & 1.17 \\
\hline\hline
\multicolumn{14}{l}{$\sigma =$ scatter (standard deviation) of the observed parameter across all observations.}\\
\multicolumn{14}{l}{$\overline{\sigma_{N}} = $ average scatter of the observed parameter measured on individual nights.}\\
\end{tabular}

\end{table*}

\subsection{Science Targets and Observations}
\label{sec:scip}
Science targets were observed as part of programmes PQ18A02, PL18A10, PL18B01 and PL19A16.  Targets were selected, primarily from ZTF, for their brightness and location in the sky to be suitable for RINGO3 observations.
The sample is composed of 4 Type II SNe, 3 Type Ia SNe, 2 1991T-like Type Ia SNe, 1 Type Ic SN, 1 Type I Superluminous SN (SLSN) and a single transient (AT2019hgp) of unknown classification.  Details of the formal discovery and classification of these transients are shown in Table \ref{tab:scip}.  We note, however, that in a number of cases the objects had ZTF observations prior to the date and time given in the formal discovery announcement.  

During the ZTF observations, difference images are generated based on the image subtraction algorithm by \citet{2016apj...830...27z} implemented in the real-time reduction pipeline \citep{2019pasp..131a8003m}.  Only ZTF alert streams that are above a $5-\sigma$ threshold will generate alerts. Using the IPAC ZTF difference imaging pipeline, we performed forced point-spread function (PSF) photometry at the location of SNe discovered by ZTF following the procedure described in \citet{2019apj...886..152y}. We applied a $4-\sigma$ threshold and inspected both the last non-detection limit and the first detection in both $g$ and $r$ bandpasses. To establish the phase of our observations, for each transient, we consider the explosion $t_{\rm exp}$ to be midway between the last ZTF non-detection and the first ZTF detection of the candidate (see Table~\ref{tab:detection}). The estimated explosion time of the Type II SNe\,2018cyg and 2018dfi are consistent with \citet{2020arxiv200809986b}. The discovery time of SN\,2018gep agrees with the first $r-$band detection reported by \citet{2019apj...887..169h}. 

\begin{figure*}
\includegraphics[]{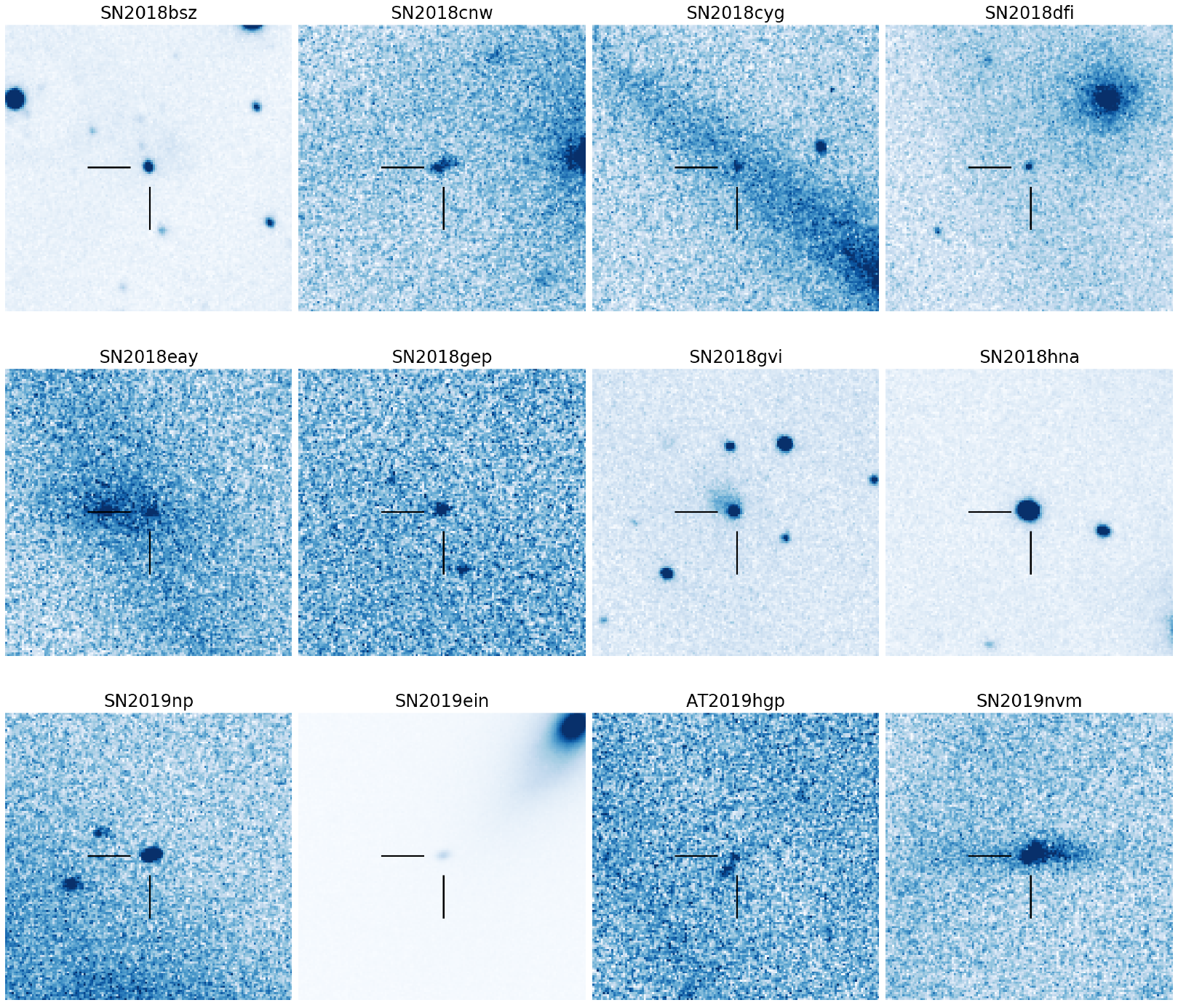}
\caption{RINGO3 $g^{\ast}$ images of the science targets.  All images are centred on the science target, have dimensions of $1\arcmin \times 1\arcmin$ and oriented with North up and East to the left.}
\label{fig:scip}
\end{figure*}

A log of the science observations is presented in Table \ref{tab:obs} and the locations of the SNe, with respect to their host galaxies, are shown on Fig. \ref{fig:scip}.  It was not possible to observe two of the science targets (SN~2018bsz and 2018hna) at early times, and these constitute outliers from the main targets of the early-time polarimetry survey.  As these were observed alongside our other targets, and using the same Liverpool Telescope programmes, we include them here for completeness.

\begin{table*}
\centering
\caption{Science targets for RINGO3 observations.
    \label{tab:scip}}
\begin{tabular}{llccccll}
\hline
Target  & Original & $\alpha_{J2000}$   & $\delta_{J2000}$  & Discovery & Redshift  & Type & Discoverer \\
        & Name     &                    &                   &  Date     &           &      &            \\

\hline
SN~2018bsz & ASASSN-18km & 16:09:39.1 & -32:03:45.6 & 2018-05-17 & 0.027 & SLSN-I \citep{2018atel11674....1a} & ASASSN$^{1}$ \citep{2018TNSTR.655....1S} \\
SN~2018cnw & ZTF~18abauprj & 16:59:05.0 & +47:14:11.2 & 2018-06-15 & 0.028 & SN Ia-91T-like \citep{2018TNSCR.833....1M} & ZTF$^{2}$ \citep{2018TNSTR.832....1F} \\
SN~2018cyg & ZTF~18abdbysy & 15:34:08.5 & +56:41:48.7 & 2018-06-30 & 0.011 & SN II \citep{2018TNSCR.939....1F} & LOSS $^{3}$\citep{2018TNSTR.907....1J} \\
SN~2018dfi & ZTF~18abffyqp & 16:50:50.1 & +45:23:52.5 & 2018-07-10 & 0.031 & SN IIb \citep{2020arxiv200809986b} &  POSS$^{4}$ \citep{2018TNSTR.957....1G}\\
SN~2018eay & ZTF~18abgmcmv & 18:16:13.1 & +55:35:27.2 & 2018-07-15 & 0.018 & SN Ia-91T-like \citep{2018TNSCR1012....1Y} & ZTF$^{2}$ \citep{2018TNSTR1005....1F} \\
SN~2018gep & ZTF~18abukavn & 16:43:48.2 & +41:02:43.4 & 2018-09-09 & 0.032 & SN Ic-BL \citep{2018TNSCR1442....1B} & ZTF$^{2}$\citep{2018TNSTR1357....1H} \\
SN~2018gvi & ZTF~18abyxwrf & 02:55:36.0 & +43:03:27.3 & 2018-09-24 & 0.021 & SN Ia \citep{2018TNSCR1487....1F} & ZTF$^{2}$ \citep{2018TNSTR1456....1F} \\
SN~2018hna & $\cdots$ & 12:26:12.1 & +58:18:50.8 & 2018-10-22 & 0.002 & SN II \citep{2018TNSCR1638....1L} & K. Itagaki \citep{2018TNSTR1614....1I} \\
SN~2019np & ZTF~19aacgslb & 10:29:22.0 & +29:30:38.3 & 2019-01-09 & 0.004 & SN Ia \citep{2019TNSCR..71....1B} &  K. Itagaki \citep{2019TNSTR..53....1I} \\
SN~2019ein &  ATLAS19ieo & 13:53:29.1 & +40:16:31.3 & 2019-05-01 & 0.008 & SN Ia \citep{2019TNSCR.701....1B} &  ATLAS$^{5}$ \citep{2019TNSTR.678....1T} \\
AT~2019hgp & ZTF~19aayejww & 15:36:12.9 & +39:44:00.6 & 2019-06-08 & $\cdots$ & $\cdots$ & ZTF$^{2}$ \citep{2019TNSTR.973....1B} \\
SN~2019nvm & ZTF~19abqhobb & 17:25:38.7 & +59:26:48.3 & 2019-08-19 & 0.019 & SN II \citep{2019TNSCR1557....1H} & ZTF $^{2}$\citep{2019TNSTR1546....1N} \\

\hline
\end{tabular}
$^{1}$ All-Sky Automated Survey for Supernovae \citep[ASAS-SN][]{2014ApJ...788...48S}; $^{2}$ Zwicky Transient Facility \citep[ZTF][]{2019pasp..131a8002b}; $^{3}$ Lick Observatory Supernova Search \citep[LOSS][]{2000aipc..522..103l}; $^{4}$ Puckett Observatory Supernova Search (POSS); $^{5}$ Asteroid Terrestrial-impact Last Alert System \citep[ATLAS;][]{2018pasp..130f4505t} 
\end{table*}

\begin{table*}
\centering
\caption{Non-detection limits and first detection epochs for the observed SNe. \label{tab:detection}}
\begin{tabular}{lccc}
\hline
Target  & Filter & Last ZTF Non-detection  & First ZTF detection \\
        &        &(MJD$^{a}$  [S/N]$^{b}$)   & (MJD$^{a}$  [S/N]$^{b}$) \\
\hline
SN\,2018bsz / ASASSN-18km  & $\cdots$  & $\cdots$     & $\cdots$   \\
SN\,2018cnw / ZTF18abauprj &    $g$    & 58282.385 [$\textless$0] &  58283.283 [5.1] \\
                           &    $r$    & 58283.329 [3.2]          &  58284.280 [19]  \\
SN\,2018cyg / ZTF18abdbysy &    $g$    & 58294.223 [$\textless$0] &  58295.205 [5.2] \\
                           &    $r$    & 58294.242 [2.1]          &  58294.257 [4.0] \\
SN\,2018dfi / ZTF18abffyqp &    $g$    & 58306.307 [$\textless$0] &  58307.214 [63]  \\
                           &    $r$    & 58306.201 [$\textless$0] &  58307.186 [44]  \\
SN\,2018eay / ZTF18abgmcmv &    $g$    & 58311.345 [3.7]$^{d}$    &  58312.350 [11]  \\
                           &    $r$    & 58311.198 [2.5]          &  58311.222 [5.5] \\
SN\,2018gep / ZTF18abukavn &    $g$    & 58369.254 [3.9]$^{d}$    &  58370.186 [19]  \\
                           &    $r$    & 58370.141 [4.0]$^{d}$    &  58370.163 [7.3] \\
SN\,2018gvi / ZTF18abyxwrf &    $g$    & 58384.319 [2.3]          &  58385.413 [6.3] \\
                           &    $r$    & 58386.328 [$\textless$0] &  58388.485 [21]  \\
SN\,2018hna / ZTF18acbwaxk & $\cdots$  & $\cdots$     & $\cdots$   \\
SN\,2019np  / ZTF19aacgslb &    $g$    & 58491.454 [$\textless$0] &  58494.483 [179] \\
                           &    $r$    & 58491.530 [1.2]          &  58492.445 [18]  \\
SN\,2019ein / ATLAS19ieo   &cyan-ATLAS & 58602.267  &  58604.474$^{c}$ \\
AT\,2019hgp / ZTF19aayejww &    $g$    & 58640.362 [$\textless$0] &  58641.201 [3.5]$^{d}$ \\
                           &    $r$    & 58640.291 [$\textless$0] &  58641.289 [4.9]  \\
%AT\,2019hgp / ZTF19aayejww &    $g$    & 58641.243 [1.7]          &  58642.303 [26]  \\
%                           &    $r$    & 58641.308 [1.8]          &  58642.242 [13]  \\
SN\,2019nvm / ZTF19abqhobb &    $g$    & 58713.218 [$\textless$0] &  58714.163 [71]  \\
                           &    $r$    & 58713.242 [1.2]          &  58714.185 [55]  \\
\hline
\end{tabular}\\
$^{a}$Modified Julian Date; $^{b}$Signal-to-noise ratio for the forced difference image PSF-fit flux measurement; \\
$^{c}$Discovered by ATLAS \citep{2019TNSTR.678....1T}; 
$^{d}$Target was marginally detected; 
\end{table*}

\begin{table*}
    \centering
        \caption{RINGO3 observations of the target SNe.  The exposure time is the total spent on the target across all 8 polaroid rotator positions.  The same exposure time is used for each of the three ($b^{\ast}/g^{\ast}/r^{\ast}$) RINGO3 channels.  
    \label{tab:obs}}
    \begin{tabular}{lrrllr}
\hline
 Date (UT)        &      MJD & Phase & Dataset         & Target           &   Exposure \\
       &       & (days)$^{\dagger}$&       &          &   Time(s) \\
\hline
  2018-06-02 23:02 & 58271.96 & 69.5 & e\_20180602\_2\_0  & SN 2018bsz  &           598 \\
  2018-06-05 22:53 & 58274.95 & 72.4 & e\_20180605\_8\_0  & SN 2018bsz  &           598 \\
  2018-06-10 22:37 & 58279.94 & 77.4 & e\_20180610\_4\_0  & SN 2018bsz  &           895 \\
  2018-06-13 22:58 & 58282.96 & 80.5 & e\_20180613\_4\_0  & SN 2018bsz  &           895 \\
  2018-06-17 22:05 & 58286.92 & 84.4 & e\_20180617\_3\_0  & SN 2018bsz  &          1196 \\
  2018-06-20 22:01 & 58289.92 & 87.4 & e\_20180620\_4\_0  & SN 2018bsz  &          1197 \\
  2018-06-24 21:35 & 58293.90 & 91.4 & e\_20180624\_1\_0  & SN 2018bsz  &          1196 \\
  2018-06-28 21:33 & 58297.90 & 95.4 & e\_20180628\_3\_0  & SN 2018bsz  &          1198 \\
 \hline
 2018-06-18 02:00 & 58287.08 & 4.2 & e\_20180617\_10\_0 & SN 2018cnw   &           948 \\
 2018-06-18 02:16 & 58287.09 & 4.3 & e\_20180617\_11\_0 & SN 2018cnw   &           948 \\
 2018-06-18 02:32 & 58287.11 & 4.3 & e\_20180617\_12\_0 & SN 2018cnw   &           948 \\
 2018-06-18 02:48 & 58287.12 & 4.3 & e\_20180617\_13\_0 & SN 2018cnw   &           945 \\
 2018-06-18 03:04 & 58287.13 & 4.3 & e\_20180617\_14\_0 & SN 2018cnw   &           945 \\
\hline
 2018-06-26 23:00 & 58295.96 & 1.7 & e\_20180626\_3\_0  & SN 2018cyg   &          1799 \\
 2018-06-26 23:30 & 58295.98 & 1.7 & e\_20180626\_4\_0  & SN 2018cyg   &          1796 \\
 2018-06-27 00:01 & 58296.00 & 1.8 & e\_20180626\_5\_0  & SN 2018cyg.  &          1796 \\
\hline
 2018-07-08 22:12 & 58307.93 & 1.2 & e\_20180708\_5\_0  & SN 2018dfi   &          1196 \\
 2018-07-08 22:32 & 58307.94 & 1.2 & e\_20180708\_6\_0  & SN 2018dfi   &          1196 \\
 2018-07-09 01:37 & 58308.07 & 1.3 & e\_20180708\_7\_0  & SN 2018dfi   &          1196 \\
 2018-07-09 01:58 & 58308.08 & 1.3 & e\_20180708\_8\_0  & SN 2018dfi   &          1196 \\
\hline
 2018-07-19 22:12 & 58318.93 & 7.7 & e\_20180719\_5\_0  & SN 2018eay   &           996 \\
 2018-07-19 22:29 & 58318.94 & 7.7 & e\_20180719\_6\_0  & SN 2018eay   &           999 \\
\hline
 2018-09-23 22:15 & 58384.93 & 14.8 & e\_20180923\_16\_0 & SN 2018gep   &           446 \\
 2018-09-23 22:23 & 58384.93 & 14.8 & e\_20180923\_17\_0 & SN 2018gep   &           446 \\
\hline
 2018-10-02 02:39 & 58393.11 & 8.2 & e\_20181001\_10\_0 & SN 2018gvi   &          1199 \\
\hline
 2019-01-10 05:07 & 58493.21 & 82.4 & e\_20190109\_4\_0  & SN 2018hna   &           476 \\
 2019-01-14 02:09 & 58497.09 & 86.3 & e\_20190113\_9\_0  & SN 2018hna   &           478 \\
 2019-01-20 01:44 & 58503.07 & 92.3 & e\_20190119\_7\_0  & SN 2018hna   &           537 \\
 2019-01-30 03:35 & 58513.15 & 102.3 & e\_20190129\_5\_0  & SN 2018hna   &           715 \\
 2019-02-04 02:54 & 58518.12 & 107.3 & e\_20190203\_5\_0  & SN 2018hna   &           957 \\
 2019-02-12 01:45 & 58526.07 & 115.3 & e\_20190211\_7\_0  & SN 2018hna   &           997 \\
\hline
 2019-01-11 03:46 & 58494.16 & 2.2 & e\_20190110\_4\_0  & SN 2019np    &           898 \\
 2019-01-12 03:48 & 58495.16 & 3.2 & e\_20190111\_4\_0  & SN 2019np    &          1197 \\
 2019-01-12 04:35 & 58495.19 & 3.2 & e\_20190111\_5\_0  & SN 2019np    &          1197 \\
 2019-01-13 01:13 & 58496.05 & 4.1 & e\_20190112\_9\_0  & SN 2019np    &          1198 \\
 2019-01-14 01:50 & 58497.08 & 5.1 & e\_20190113\_8\_0  & SN 2019np    &           998 \\
 2019-01-15 02:34 & 58498.11 & 6.1 & e\_20190114\_20\_0 & SN 2019np    &           717 \\
 2019-01-16 06:40 & 58499.28 & 7.3 & e\_20190115\_10\_0 & SN 2019np    &           598 \\
 2019-01-20 01:35 & 58503.07 & 11.1 & e\_20190119\_6\_0  & SN 2019np    &           417 \\
 2019-01-21 01:40 & 58504.07 & 12.1 & e\_20190120\_3\_0  & SN 2019np    &           417 \\
 2019-01-23 01:33 & 58506.06 & 14.1 & e\_20190122\_9\_0  & SN 2019np    &           717 \\
 2019-01-26 04:55 & 58509.21 & 17.2 & e\_20190125\_3\_0  & SN 2019np    &           717 \\
 2019-02-07 01:15 & 58521.05 & 29.1 & e\_20190206\_9\_0  & SN 2019np    &           996 \\
 2019-02-25 00:11 & 58539.01 & 47.0 & e\_20190224\_3\_0  & SN 2019np    &          1196 \\
\hline
 \end{tabular}
 \\
 $^{\dagger}$ Relative to calculated explosion epoch (see Section \ref{sec:scip} and Table \ref{tab:detection}).
\end{table*}

\begin{table*}
    \centering
        \contcaption{RINGO3 observations of the target SNe.  The exposure time is the total spent on the target across all 8 polaroid rotator positions.  The same exposure time is used for each of the three wavelength channels.
    \label{tab:obs:cont}}
    \begin{tabular}{lrrllr}
\hline
 Date (UT)        &      MJD & Phase  & Dataset         & Target           &   Exposure \\
    &       &   (days) &    &         &  Time(s) \\
\hline
 2019-05-03 21:27 & 58606.89 & 3.5 & e\_20190503\_3\_0  & SN 2019ein       &          1499 \\
 2019-05-04 21:46 & 58607.91 & 4.5 & e\_20190504\_3\_0  & SN 2019ein        &          1497 \\
 2019-05-04 22:13 & 58607.93 & 4.6 & e\_20190504\_4\_0  & SN 2019ein        &          1497 \\
 2019-05-05 21:32 & 58608.90 & 5.5 & e\_20190505\_3\_0  & SN 2019ein         &          1497 \\
 2019-05-06 21:56 & 58609.91 & 6.5 & e\_20190506\_3\_0  & SN 2019ein        &          1496 \\
 2019-05-07 22:47 & 58610.95 & 7.6 & e\_20190507\_3\_0  & SN 2019ein         &           896 \\
 2019-05-09 22:47 & 58612.95 & 9.6 & e\_20190509\_3\_0  & SN 2019ein       &           599 \\
 2019-05-11 00:04 & 58614.00 & 10.6 & e\_20190510\_3\_0  & SN 2019ein         &           598 \\
 2019-05-11 23:03 & 58614.96 & 11.6 & e\_20190511\_5\_0  & SN 2019ein         &           598 \\
 2019-05-14 00:45 & 58617.03 & 13.7 & e\_20190513\_26\_0 & SN 2019ein       &           598 \\
  2019-05-16 23:03 & 58619.96 & 16.6 & e\_20190516\_4\_0  & SN 2019ein        &           599 \\
  2019-05-19 21:40 & 58622.90 & 19.5 & e\_20190519\_1\_0  & SN 2019ein        &           598 \\
  2019-05-22 22:30 & 58625.94 & 22.6 & e\_20190522\_1\_0  & SN 2019ein         &           598 \\
  2019-05-27 22:20 & 58630.93 & 27.6 & e\_20190527\_3\_0  & SN 2019ein         &           596 \\
  2019-06-04 23:10 & 58638.97 & 35.6 & e\_20190604\_3\_0  & SN 2019ein          &           598 \\
  2019-06-24 23:24 & 58658.98 & 55.6 & e\_20190624\_8\_0  & SN 2019ein        &          1196 \\
  2019-07-05 22:23 & 58669.93 & 66.6 & e\_20190705\_3\_0  & SN 2019ein        &          1797 \\
  2019-07-26 21:09 & 58690.88 & 87.5 & e\_20190726\_3\_0  & SN 2019ein &          1798 \\
\hline
 2019-06-09 22:56 & 58643.96 & 3.2 & e\_20190609\_3\_0 & AT 2019hgp &          1797 \\
\hline
 2019-08-19 21:32 & 58714.90 & 1.2 & e\_20190819\_4\_0 & SN 2019nvm &          1797 \\
 \hline
 \end{tabular}
\end{table*}

%%%%%%%%%%%%%%%%%%%%%%%%%%%%%%%%%%%%%%%%%%%%%%%
%RESULTS AND ANALYSIS
%RESULTS AND ANALYSIS
%RESULTS AND ANALYSIS
%%%%%%%%%%%%%%%%%%%%%%%%%%%%%%%%%%%%%%%%%%%%%%%
\section{Results \& Analysis}
\label{sec:res}
%
%SN2018bsz
%
\subsection{SN~2018bsz}
\label{sec:res:18bsz}
\begin{figure}
    \centering
    \includegraphics[width=8.5cm]{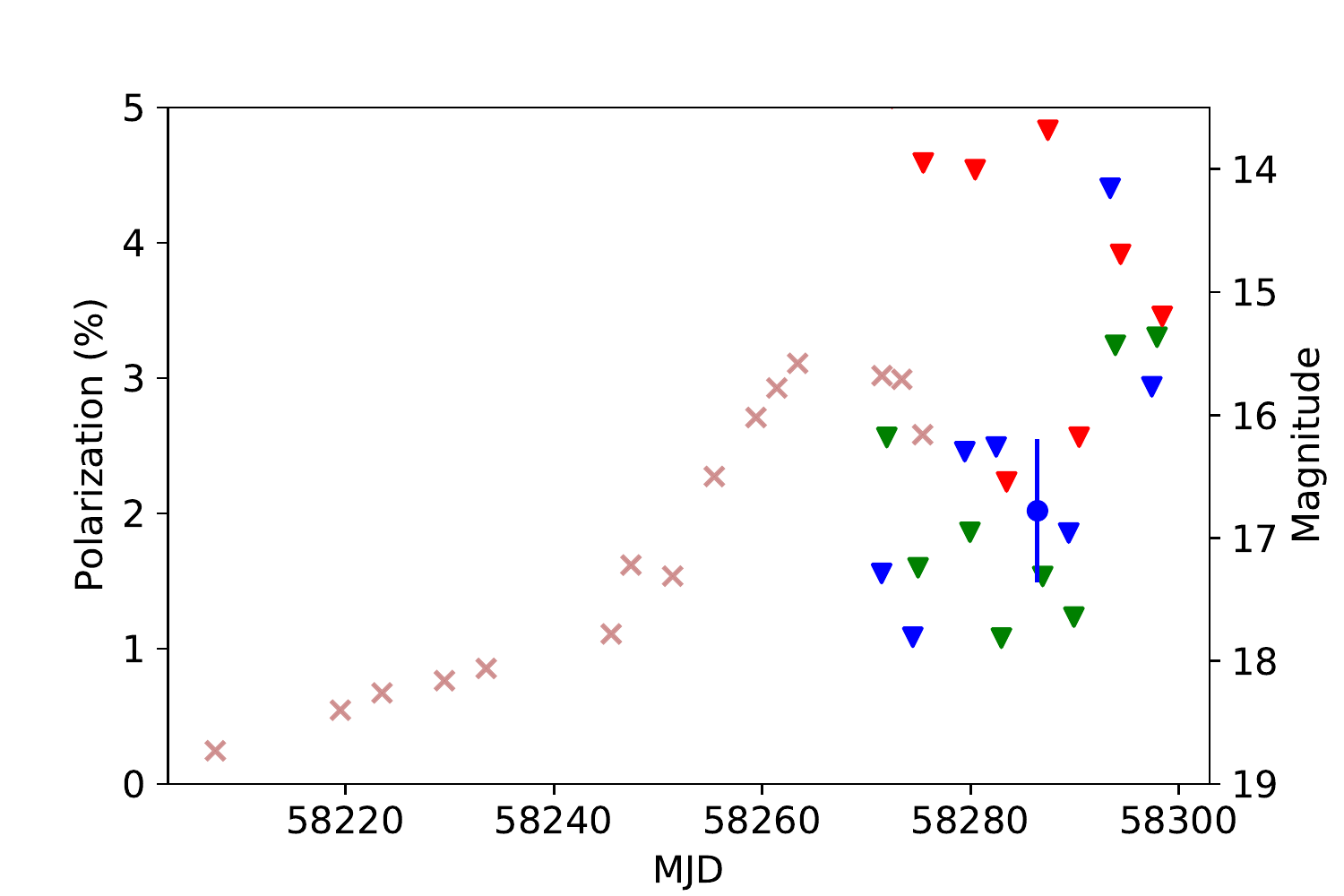}
    \caption{The evolution of the polarization of SN~2018bsz.  The measurements are colour coded according to the RINGO3 channels $b^{\ast}$ (blue), $g^{\ast}$ (green) and $r^{\ast}$ (red) and are composed of detections ($\bullet$) and upper limits ($\blacktriangledown$).  Also shown is the ATLAS $o$-band photometry ($\times$) reported by \citet{2018a&a...620a..67a}.}
    \label{fig:res:18bsz}
\end{figure}

Discovered by ASAS-SN on 2018 May 17 \citep{2018TNSTR.655....1S}, it was temporarily classified as a young Type II SN \citep{2018TNSCR.679....1H}.  It was later reclassified as a superluminous supernova \citep[for a review see ][]{2019ara&a..57..305g}, albeit a lower luminosity example \citep{2018a&a...620a..67a}.  Our observations commenced 4 days after the photometric light curve maximum or 69 days after the explosion date proposed by \citet{2018a&a...620a..67a}.  The polarization measurements for SN~2018bsz are presented in Table \ref{tab:res:18bsz} and the time evolution, with respect to the photometric light curve, is shown on Fig. \ref{fig:res:18bsz}.  We derive limits on the polarization, strictest in the blue, at the general level of $<1  - 2\%$.  We do, however, make one single detection of $p(b^{\ast}) = 2 \pm 0.5$\%  at 11.4 days after maximum (or MJD58267.5).

\begin{table}
    \centering
    \caption{RINGO3 polarization measurements of SN~2018bsz. \label{tab:res:18bsz}}
    \begin{tabular}{lrccc}
\hline
Epoch  &  Phase &  $p(b^{\ast})$ &    $p(g^{\ast})$ &    $p(r^{\ast})$   \\
(MJD)  & (days) &  (\%)  & (\%) & (\%) \\
\hline
58271.96 & 69.5 &$<1.56$ & $<2.56$ & $<5.07$\\
58274.95 & 72.4 &$<1.08$ & $<1.60$ & $<4.59$\\
58279.94 & 77.4 &$<2.46$ & $<1.86$ & $<4.54$\\
58282.96 & 80.5 &$<2.49$ & $<1.08$ & $<2.23$\\
58286.92 & 84.4 &$2.02\pm0.53$ & $<1.53$ & $<4.83$\\
58289.92 & 87.4 &$<1.85$ & $<1.23$ & $<2.56$\\
\hline
    \end{tabular}

\end{table}

%
%SN 2018cnw
%
\subsection{SN~2018cnw}
\label{sec:res:18cnw}
\citet{2018TNSCR.833....1M} classified SN~2018cnw as being a ``91T-like" Type Ia SN.  Five sets of observations were acquired in the course of a single night (around Modfied Julian Date [MJD] 58287.0).  The observation of the SN was subject to poor seeing of $\sim 3.2\arcsec$, with the point spread function appearing obviously elongated.  We note that it was not possible, under these conditions, to detect the SN in the $r^{\ast}$ observations.  For the other filters, we derived upper limits on the polarization of $b^{\ast} < 2.0\%$ and $g^{\ast} < 0.9\%$.

%
%SN2018cyg
%
\subsection{SN~2018cyg}
\label{sec:res:18cyg}
SN~2018cyg was observed three times in a single night (MJD58296.0).  The SN was marginally detected (S/N = 4.0) in ZTF $r-$band observation on MJD58294.257, followed by the $g-$ and $r-$band detections at MJD58295.205 (S/N = 5.2) and MJD 58295.246 (S/N = 16.0), respectively.  We consider the RINGO3 observation was, therefore, conducted at $\sim 0.8$ to $1.8$ days after the first detection with ZTF, or $\approx$2 days after the explosion. (see Table~\ref{tab:detection}).  \citet{2018TNSCR.939....1F} later classified it as a Type II SN, with a short plateau. Given the faintness of the SN at this epoch and poor observing conditions, we could not establish a photometric detection of SN~2018cyg in the $b^{\ast}$-band and were only able to place limits on the degree of polarization of $g^{\ast} < 15\%$ and $r^{\ast} < 22.0\%$.

%
%SN2018dfi
%
\subsection{SN~2018dfi}
\label{sec:res:18dfi}
Two sets of observations of SN~2018dfi, consisting of two separate exposures each, were conducted on a single night (MJD58308.0).  The RINGO3 observations commenced 0.7 days after the first detection by ZTF, which we estimate to correspond to 1.2 days post-explosion. The SN was classified (at 4.5 days post-explosion) as a Type II SN \citep{2018TNSCR.974....1H}, which was further refined to being Type IIb \citep{2020arxiv200809986b}.  Combining the Stokes parameters determined for all four exposures, we derive limits on the degree of polarization of SN~2018dfi of  $p(b^{\ast}) < 2.3\%$, $p(g^{\ast}) < 6.8\%$ and  and $p(r^{\ast}) < 4\%$.

%
%SN2018eay
%
\subsection{SN~2018eay}
\label{sec:res:18eay}
Forced PSF photometry of ZTF images at the location of SN\,2018eay shows that the S/N measured in both $g$ (S/N=3.7 at MJD 58311.345) and $r$ (S/N=5.5 at MJD 58311.222) bands are obviously higher compared to the previous non-detections. The S/N derived based on an $r-$band image obtained earlier during the same night (MJD 58311.198) yields 2.5. Therefore, we adopt an explosion epoch at MJD 58311.2 for SN\,2018eay, indicating that the first RINGO3 imaging polarimetry was acquired 7.7 days post the SN explosion (the night of MJD 58319.0).  Similarly to SN~2018cnw, SN~2018eay was classified as a ``91T-like" Type Ia SN \citep{2018TNSCR1012....1Y}.  In both sets of observations, the SN was only weakly detected in all three channels and it was only possible to derive limits on the degree of polarization of $p(b^{\ast}) < 2.5\%$, $p(g^{\ast}) < 6.1\%$ and $p(r^{\ast}) < 7.5\%$.

%
%SN2018gep
%
\subsection{SN~2018gep}
\label{sec:res:18gep}
Two sets of observations of SN~2018gep were conducted, in sequence, on the night of MJD58385.0, corresponding to 14.8 days post-explosion.  We note that SN~2018gep was discovered very close to the moment of explosion, with the last ZTF non-detection of the transient occurring only 0.02 days before the first detection \citep{2018TNSTR1357....1H}.
They conducted a second-order polynomial fit to the first three days of the $g$-band flux and defined $t_{0}$ at 58370.146 as the time at which the flux of SN\,2018gep is zero. In fact, the transient exhibits pre-explosion emission extended $\approx$ a week prior to the rapid rise in the light curve (see Figure 7 of \citealp{2019apj...887..169h}).  An observation of SN~2018gep at 10.1 days yielded a classification for SN~2018gep as a broad-lined Type Ic supernova at around maximum light \citep{2018TNSCR1442....1B}; although \citet{2020arxiv200804321p} suggest the fast rise-time ($<6.2\,\mathrm{days}$) may imply that SN2018gep may be more closely related to the family of Fast Blue Optical Transients \citep{2014apj...794...23d}.  From our RINGO3 observations we do not detect any significant polarization for SN~2018gep, instead deriving polarization limits of $p(b^{\ast}) < 1.6\%$ and $p(g^{\ast}) <7.0\%$ and $p(r^{\ast}) < 5.1\%$.

%
%SN2018gvi
%
\subsection{SN~2018gvi}
\label{sec:res:18gvi}
We acquired a single observation of SN~2018gvi at 8.2 days post-explosion, or 7.7 days after the first detection by ZTF.  The SN had earlier been classified by \citet{2018TNSCR1487....1F} as a Type Ia SN.  We derive upper limits on the polarization of SN~2018gvi in all three channels to levels of $p(b^{\ast}) < 1.6\%$, $p(g^{\ast}) < 2.0\%$ and $p(r^{\ast}) < 5.0\%$.

%
%SN2018hna
%
\subsection{SN~2018hna}
\label{sec:res:18hna}
\begin{figure}
    \centering
    \includegraphics[width=8.5cm]{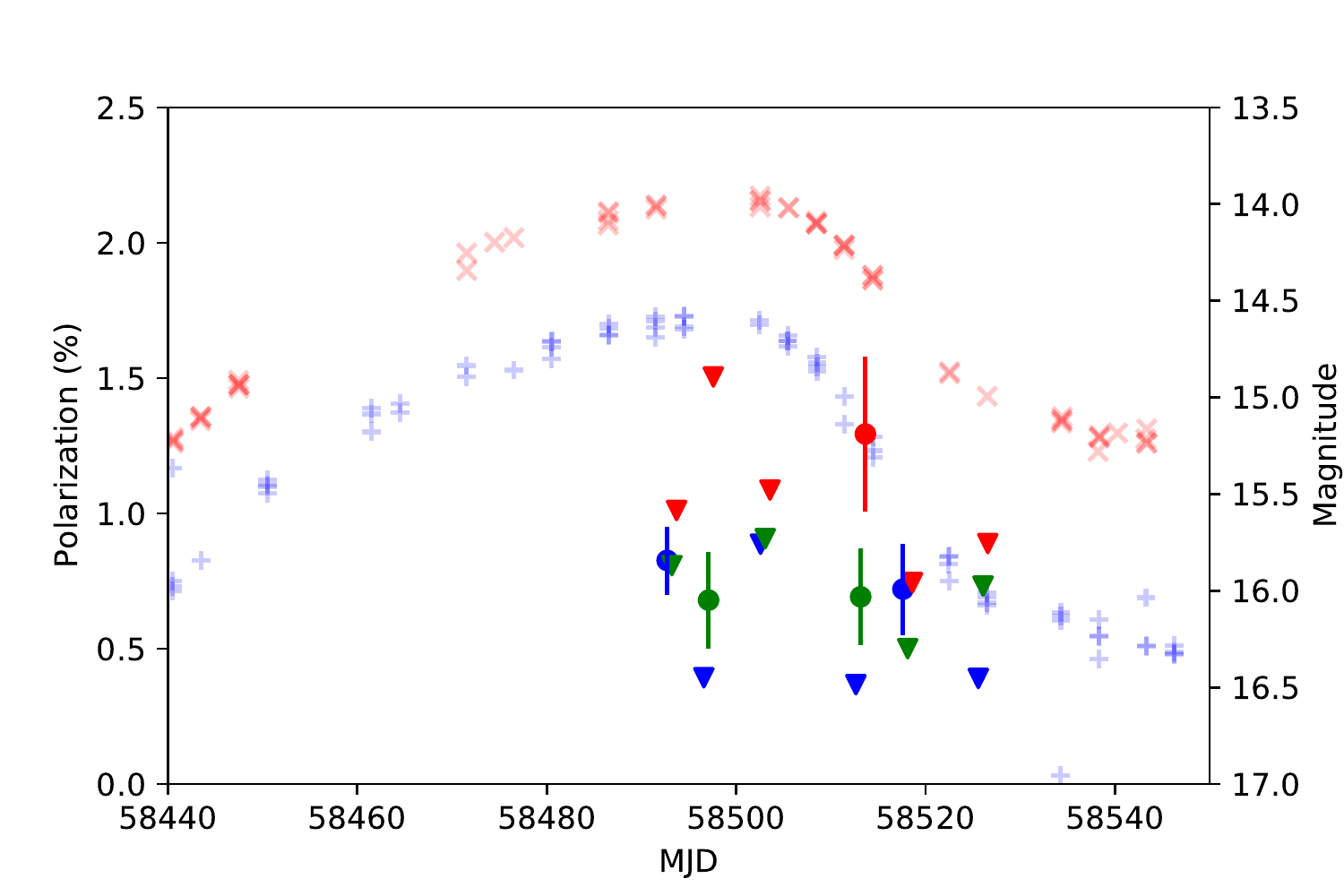}
    \caption{The evolution of the polarization of SN~2018hna (see Table \ref{tab:res:18hna}) using the same plotting scheme as Fig. \ref{fig:res:18bsz}.  Also shown is ZTF $g^{\prime}$ ($+$) and $r^{\prime}$ ($\times$) photometry.}
    \label{fig:res:18hna}
\end{figure}

SN~2018hna was observed with RINGO3 at 6 separate epochs, beginning $\sim 5$ days before the $V$-band maximum.  Given the long rise time to maximum light, the observations commenced $\sim 82$ days post-explosion as shown in Fig. \ref{fig:res:18hna}\footnote{\url{https://lasair.roe.ac.uk/object/ZTF18acbwaxk/}} \citep{2019apj...882l..15s}.  The polarization measurements at each of the six epochs are listed in Table \ref{tab:res:18hna}.  Overall, we constrain the polarization of SN~2018hna at around maximum light to be $<1.5\%$.  In the $b^{\ast}$ and $g^{\ast}$ bands, we make four separate detections of significant polarization $p \sim 0.5 - 1.0\%$, whilst at a single epoch we find $p(r^{\ast}) = 1.3\pm0.3\%$.  From Fig. \ref{fig:res:18hna}, it is clear that any polarization associated with SN~2018hna, despite its brightness ($m_{\mathrm{max}}(r) \sim 14.0\,\mathrm{mags}$), is close to the systematic floor of the RINGO3 instrument.

\begin{table}
    \centering
    \caption{RINGO3 polarization measurements of SN~2018hna \label{tab:res:18hna}}
    \begin{tabular}{lrccc}
\hline
Epoch  & Phase &  $p(b^{\ast})$ &    $p(g^{\ast})$ &    $p(r^{\ast})$   \\
(MJD)  &  (days) & (\%)  & (\%) & (\%) \\
\hline
58493.21 & 82.4 &$0.83^{+0.12}_{-0.13}$ & $<0.81$ & $<1.01$\\
58497.09 & 86.3 &$<0.39$ & $0.68 \pm 0.18$ & $<1.50$\\
58503.07 & 92.3 &$<0.89$ & $<0.91$ & $<1.09$\\
58513.15 & 102.3&$<0.37$ & $0.69\pm0.18$ & $1.29^{+0.28}_{-0.29}$\\
58518.12 & 107.3&$0.72 \pm0.17$ & $<0.50$ & $<0.74$\\
58526.07 & 115.3&$<0.39$ & $<0.73$ & $<0.89$\\
\hline
    \end{tabular}

\end{table}
%
%SN2019np
%
\subsection{SN~2019np}
\label{sec:res:19np}
\begin{figure}
    \centering
    \includegraphics[width=8.5cm]{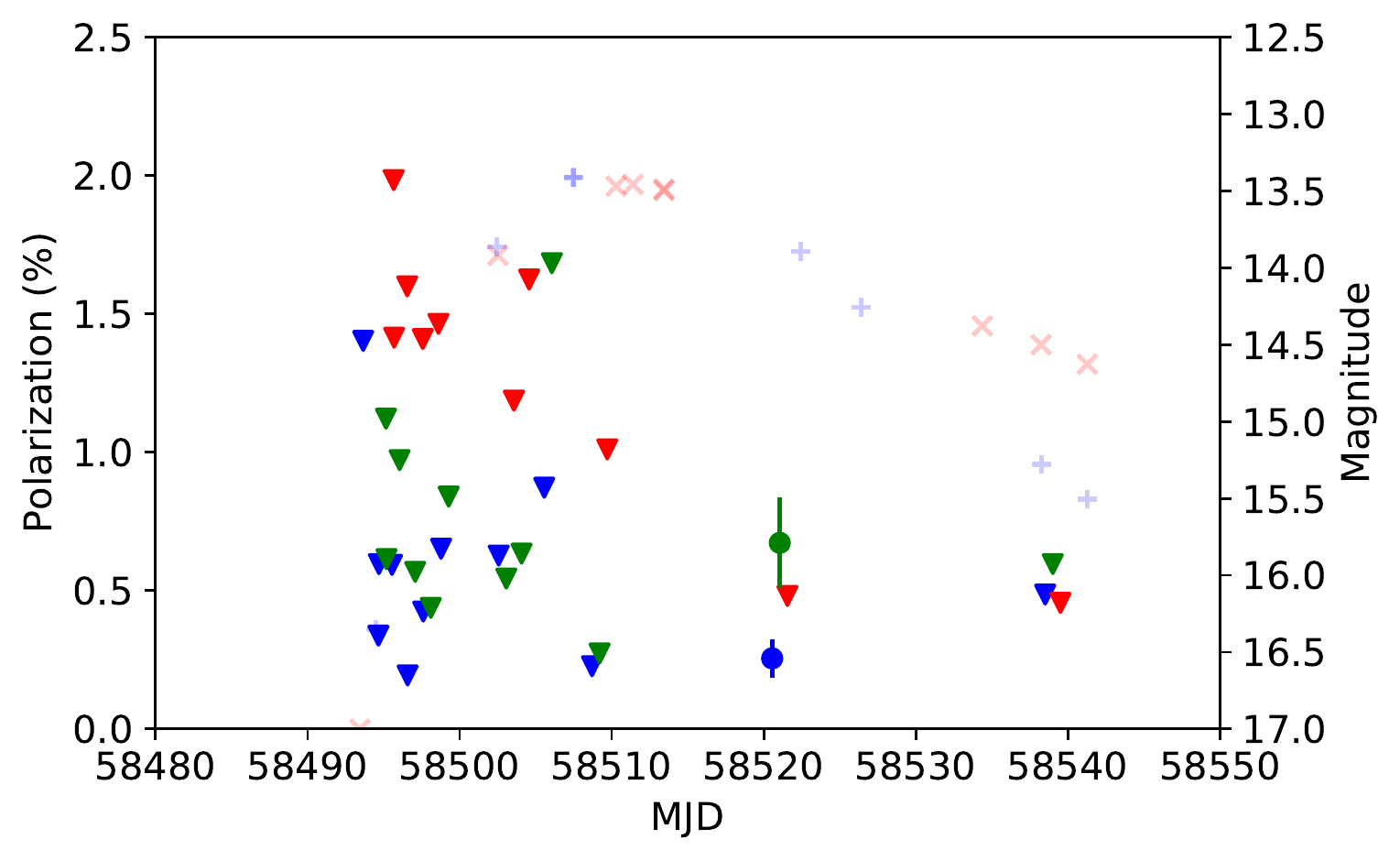}
    \caption{The evolution of the polarization of SN~2019np (see Table \ref{tab:res:19np}) using the same plotting scheme as Fig. \ref{fig:res:18bsz}.  Also shown is ZTF $g^{\prime}$ ($+$) and $r^{\prime}$ ($\times$) photometry.}
    \label{fig:res:19np}
\end{figure}

Polarimetric followup of SN~2019np commenced 2.2 days after explosion, or 1.7 days after the first detection by ZTF.  In total there were 13 separate observations of the SN up to 47 days post-explosion.  The polarization measurements for SN~2019np are presented in Table \ref{tab:res:19np} and shown in Figure \ref{fig:res:19np} (alongside ZTF photometry\footnote{\url{https://lasair.roe.ac.uk/object/ZTF19aacgslb/}}).  The SN was discovered on the rise up to maximum light and polarization constraints, in particular in the blue, limit the polarization across the optical wavelength range to $<2.0\%$.  At a later epoch, 28 days after discovery and 10 days after maximum, we detect significant polarization at the level of $p(b^{\ast}) = 0.26\pm0.07\% $ and  $p(g^{\ast})= 0.67\pm0.16\%$ consistent with the earlier limits on the polarization and the general level of polarization of this SN being low.

\begin{table}
    \centering
    \caption{RINGO3 polarization measurements of SN~2019np \label{tab:res:19np}}
    \begin{tabular}{lrccc}
\hline
Epoch  & Phase &  $p(b^{\ast})$ &    $p(g^{\ast})$ &    $p(r^{\ast})$   \\
(MJD)  & (days) & (\%)  & (\%) & (\%) \\
\hline
58494.16 & 2.2 &$<1.40$ & $<3.01$ & $<4.70$ \\
58495.16 & 3.2 &$<0.34$ & $<1.12$ & $<1.98$ \\
58495.19 & 3.2 &$<0.60$ & $<0.61$ & $<1.41$ \\
58496.05 & 4.1 &$<0.59$ & $<0.97$ & $<1.60$ \\
58497.08 & 5.1 &$<0.19$ & $<0.57$ & $<1.41$ \\
58498.11 & 6.1 &$<0.42$ & $<0.44$ & $<1.46$ \\
58499.28 & 7.3 &$<0.65$ & $<0.84$ & $<3.87$ \\
58503.07 & 11.1&$<0.63$ & $<0.54$ & $<1.19$ \\
58504.07 & 12.1&$<2.40$ & $<0.63$ & $<1.62$ \\
58506.06 & 14.1&$<0.87$ & $<1.68$ & $<2.40$ \\
58509.21 & 17.2&$<0.23$ & $<0.27$ & $<1.01$ \\
58521.05 & 29.1&$0.26\pm0.07$ & $0.67\pm0.16$ & $<0.48$ \\
58539.01 & 47.0&$<0.49$ & $<0.60$ & $<0.46$ \\
\hline
    \end{tabular}
\end{table}

%
%SN2019ein
%
\subsection{SN~2019ein}
\label{sec:res:19ein}
\begin{figure}
    \centering
    \includegraphics[width=8.5cm]{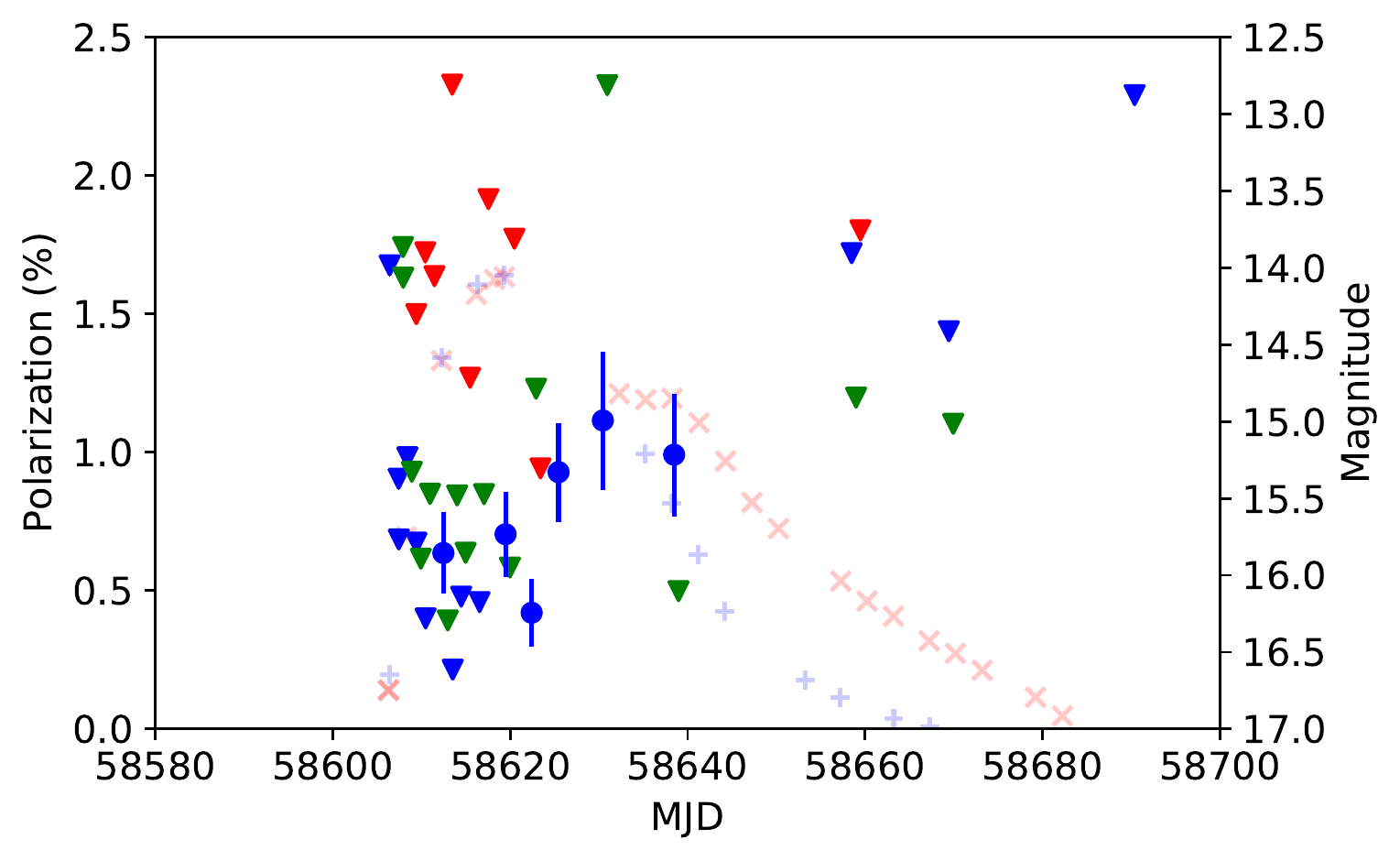}
    \caption{The evolution of the polarization of SN~2019ein (see Table \ref{tab:res:19ein}) using the same plotting scheme as Fig. \ref{fig:res:18bsz}.  Also shown is the ZTF $g^{\prime}$ ($+$) and $r^{\prime}$ ($\times$) photometry.}
    \label{fig:res:19ein}
\end{figure}

RINGO3 observations of SN~2019ein started 3.5 days post-explosion or 2.4 days after the first ZTF detection.  The Type Ia SN \citep{2019TNSCR.701....1B} was heavily observed in the rise to maximum, and in general we were only able to assess upper limits on the degree of polarization, however we did measure significant levels of polarization in the $b^{\ast}$-band (see Table \ref{tab:res:19ein}) that appear to increase with time around the period of the second light curve maximum that was observable at redder wavelengths (see Figure \ref{fig:res:19ein}, in conjunction with ZTF photometry \footnote{\url{https://lasair.roe.ac.uk/object/ZTF19aatlmbo/}}). 
\begin{table}
    \centering
    \caption{RINGO3 polarization measurements of SN~2019ein \label{tab:res:19ein}}
    \begin{tabular}{lrccc}
\hline
Epoch  & Phase &   $p(b^{\ast})$ &    $p(g^{\ast})$ &    $p(r^{\ast})$   \\
(MJD)  & (days) &  (\%)  & (\%) & (\%) \\
\hline
58606.89 & 3.5 &$<1.68$ & $<2.61$ & $<8.55$ \\
58607.91 & 4.5 &$<0.90$ & $<1.74$ & $<5.31$ \\
58607.93 & 4.6 &$<0.68$ & $<1.63$ & $<3.99$ \\
58608.90 & 5.5 &$<0.98$ & $<0.93$ & $<1.50$ \\
58609.91 & 6.5 &$<0.67$ & $<0.62$ & $<1.72$ \\
58610.95 & 7.6 &$<0.40$ & $<0.85$ & $<1.64$ \\
58612.95 & 9.6 &$0.64 \pm 0.15$  & $<0.39$ & $<2.33$ \\
58614.00 & 10.6 &$<0.21$ & $<0.84$ & $<2.54$ \\
58614.96 & 11.6 &$<0.48$ & $<0.64$ & $<1.27$ \\
58617.03 & 13.7 &$<0.46$ & $<0.85$ & $<1.92$ \\
58619.96 & 16.6 &$0.70 \pm 0.15$ & $<0.58$ & $<1.77$ \\
58622.90 & 19.5 &$0.42\pm0.12$ & $<1.23$ & $<0.94$ \\
58625.94 & 22.6 &$0.93\pm 0.18$ & $<3.15$ & $<6.83$ \\
58630.93 & 27.6 &$1.12 \pm 0.25$ & $<2.33$ & $<4.64$ \\
58638.97 & 35.6 &$0.99 \pm 0.22$ & $<0.50$ & $<8.31$ \\
58658.98 & 55.6 &$<1.72$ & $<1.20$ & $<1.80$ \\
58669.93 & 66.6 &$<1.44$ & $<1.10$ & $<2.75$ \\
58690.88 & 87.5 &$<2.29$ & $<2.74$ & $<11.17$ \\
\hline
\end{tabular}
\end{table}

%
%AT2019hgp
%
\subsection{AT~2019hgp}
\label{sec:res:19hgp}
AT2019hgp was discovered as a young transient at MJD 58642.242 \citep{2019tnsan..30....1b}. An $r-$band detection at MJD 58641.289 (S/N$=$4.9) and a $g-$band signal at MJD 58641.201 (S/N$=$3.5) were recovered by forced PSF photometry. An early spectrum obtained at MJD 58642.43 revealed emission lines of highly ionized species \citep{2019tnsan..30....1b}, suggesting that the candidate was a young and hot transient.
We note, however, that no further specific classification of this transient has been recorded.  Our RINGO3 observation was conducted 2.8 days after discovery or $\sim$3.2 days post-explosion.  The transient was only photometrically detected at significant levels in the $b^{\ast}$ observation, for which we derive an upper limit on the polarization of $p(b^{\ast}) < 5.8\%$.
%
%SN2019nvm
%
\subsection{SN~2019nvm}
\label{sec:res:19nvm}
The single RINGO3 observation of SN~2019nvm commenced 0.7 days after the SN was first detected by ZTF, or $\sim 1.2$ days post-explosion. A spectrum, acquired $\sim 16$ hours after the RINGO3 observation revealed a young Type II SN showing ``flash features'' \citep{2017natph..13..510y} of He {\sc ii} $\lambda 4686$ and N {\sc iv} $\lambda 4537$ \citep{2019TNSCR1557....1H}.  This suggests that SN~2019nvm was discovered very soon after explosion, potentially making this RINGO3 polarimetric observation the earliest ever acquired for a Type II SN.  The observations were, however, conducted under poor seeing conditions ($\approx 2.7\arcsec$), with an elongated point spread function possibly indicating the effect of wind on the telescope.  SN~2019nvm is located close to the nucleus in the edge-on galaxy UGC 10858 and, given the seeing conditions, it was not possible to accurately separate the SN and the host galaxy.  For all three channels, we do not significantly detect a polarization signal, with upper limits on the degree of polarization of $p(b^{\ast}) < 1.5\% $, $p(g^{\ast}) < 2.7\%$ and $p(r^{\ast}) < 2.2\%$.

%%%%%%%%%%%%%%%%%%%%%%%%%%%%%%%
% DISCUSSION
% DISCUSSION
% DISCUSSION
%%%%%%%%%%%%%%%%%%%%%%%%%%%%%%%
\section{Discussion \& Conclusions}
\label{sec:disc}

\begin{figure*}
\includegraphics[width=15cm]{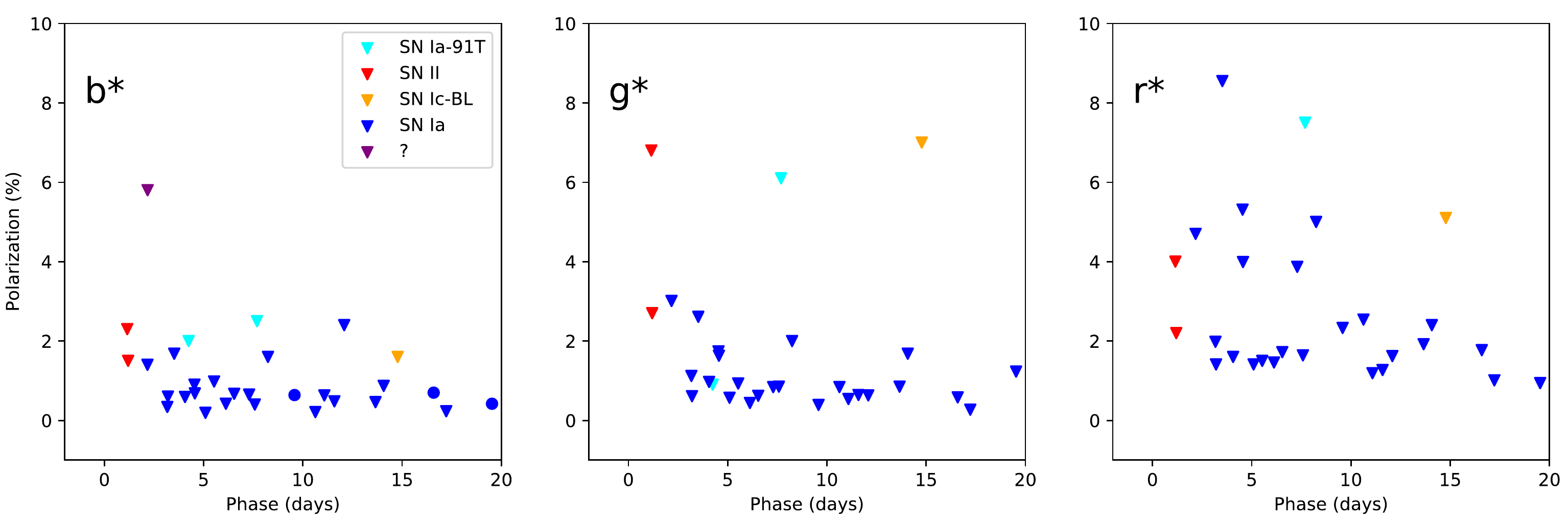}
\caption{All polarization measurements for all 10 transients with early-time observations for the first 20 days after the assumed epoch of explosion. Detections and limits on the polarization are colour-coded according to the type of the transient (according to the legend in the left panel).}
\label{fig:disc:all}
\end{figure*}

For the SNe observed at early times we have, in general, only been able to place upper limits on the possible polarization.  The evolution of the polarization constraints for all the early time observations, within 20 days of explosion, are shown in Figure \ref{fig:disc:all}.  The degree of the constraint on the early time polarization is limited by two key factors: the brightness of the SN and the relatively high level of the systematic floor of the RINGO3 instrument \citep{2016mnras.458..759s}.  In general, from our observations, the limits on the instrumental polarization means that the upper limits on the polarization are relatively high $\sim 1\%$.  As time progresses, we note that the polarization limits do become better, and this is correlated with increased levels of signal-to-noise as the target SNe rise to maximum light.  In addition, we find that the throughput in the $b^{\ast}$ channel is the best of the three RINGO3 channels and, from Fig. \ref{fig:disc:all}, the systematic floor for RINGO3 does increase towards the red.  The $r^{\ast}$-band polarization limits are less constraining by a factor of $\sim 2$.

The constraints on the level of polarization in the $b^{\ast}$ and $g^{\ast}$ bands are around $<1\%$ for Type Ia SNe.  From \citet{1991a&a...246..481h}, we can place a constraint on the axial symmetry of the ejecta of Type Ia SNe at early times (assuming a spheroidal configuration) of $>0.9$ at $t \sim 3\,\mathrm{days}$.  These limits are consistent with the earliest spectropolarimetric observation for a Type Ia SN (2018gv) which exhibited a low-level of continuum polarization ($\lesssim 0.2\%$) at $\sim 5$ days post-explosion \citep{2020apj...902...46y}.   These limits imply that Type Ia SNe generally appear almost, if not completely, spherically symmetric at the earliest times.  For the small number of CCSNe in this sample, the earliest constraints available could allow for the presence of significant asymmetries in the first few days; however, for these SNe the lack of subsequent follow-up observations at later epochs (see Fig. \ref{fig:disc:all}), as the SNe get brighter, means it is not possible to establish a baseline level of polarization (and some measure of the constant interstellar polarization [ISP] components) at maximum light as was done, for example, with SN~2019ein.   For the Type II SNe, the early limits, in particular in the $b^{\ast}$ band, constrain the axial ratio to be $>0.65$ within $\sim 1$ day of explosion.  We note that these figures do not include corrections for the ISP, a constant source of polarization, across all epochs, arising from intervening dust along the line of sight that is independent of the evolution of the polarization of the SNe themselves \citep{1973iaus...52..145s}.  We would expect, more generally, that the constraints on the intrinsic polarization of the SNe would be lower, if a correction of the ISP could be applied.

Given the limited number of significant detections of polarization across the whole sample, the Type Ia SN 2019ein stands out as having a series of detections in the $b^{\ast}$-band with a possible increasing degree of polarization (see Fig. \ref{fig:res:19ein}), reaching a maximum at around the second light curve peak observed in the ZTF $r^{\prime}$-band.  SN 2019ein has the highest recorded expansion velocities for a Type Ia SN at early times \citep{2020apj...897..159p, 2020apj...893..143k}.  SN~2019ein also exhibited blue-shifted line profiles in early spectra (at -14 days relative to B-band maximum light), in particular for the strong Si II feature.  The expansion velocities decreased as the SN approached the B-band lightcurve maximum.  \citet{2020apj...897..159p} interpreted the peculiar high expansion velocities as indicative of an asymmetric off-centre explosion, in which intermediate mass elements were mixed into the highest velocity portions of the ejecta.  The largest measured polarization of $p(b^{\ast}) = 1.12\pm0.25\%$ occurred $\sim 11$ days after the B-band lightcurve maximum.  This measurement supports the presence of significant asymmetries in SN~2019ein, as suspected from the earlier spectroscopic observations, however the time delay may indicate that the peculiar expansion velocities and the polarization are measuring the asymmetry in different ways.  As we are conducting broad-band polarimetry, however, we are not sensitive to the key diagnostic feature of Type Ia SN asymmetries, namely the polarization profile of the strong Si II line \citep{2007sci...315..212w, 2010apj...725l.167m, 2019mnras.490..578c}.  The lower polarization detections for SN~2019np ($<1\%$; see Fig. \ref{fig:res:19np}) are more generally consistent with the low levels of intrinsic continuum polarization usually seen for Type Ia SNe \citep{2008ara&a..46..433w}.

SN2018bsz joins a small, but growing collection of SLSNe with multi-epoch polarimetric observations \citep{2015apj...815l..10l, 2016apj...831...79i,
2017apj...837l..14l, 2019mnras.482.4057m}.  In general, Type I SLSNe are noted for having low levels of polarization, however SN~2015bn exhibited a significant rise in polarization after $\sim 20\,\mathrm{days}$ post-maximum \citep{2017apj...837l..14l}, with the evolution from the pre-maximum to post-maximum state clearly evident in spectropolarimetric observations \citep{2016apj...831...79i}.  Our detection of significant polarization for SN~2018bsz also occurs $\sim 20\,\mathrm{days}$ (rest-frame) after the time of the V-band light curve maximum \citep{2018a&a...620a..67a}.  \citet{2016apj...831...79i} and \citet{2017apj...837l..14l} both explained the behaviour of SN~2015bn as being due to a fundamental change in the asymmetry of the ejecta, with early time emission arising from an almost spherical outer layer, whilst at later times the emission arise from a more aspherical interior (giving rise to the increase in polarization with time).  Due to the limitations of RINGO3, we were only able to obtain single detection of polarization at $\sim 2\%$.  At a similar epoch in the evolution of SN~2015bn, \citeauthor{2017apj...837l..14l} measured $\sim 1\%$ (although the degree of polarization later rose to $\sim 1.54\%$ by $\sim 46\,\mathrm{days}$).  It is interesting to note that \citeauthor{2016apj...831...79i} and \citeauthor{2017apj...837l..14l} showed that, although the interior of 2015bn was more aspherical than the outer layers, the orientation of the asymmetry (in the plane of the sky) was the same at both early (before the light curve maximum) and later times (after the light curve maximum).   Future dense time series of polarimetric observations of Type I SLSNe will be able to confirm if the rise in polarization, coupled with the stability of the polarization angle, is a common feature for this class of SLSN.

SN~2018hna appeared, photometrically and spectroscopically, similar to SN~1987A, with the lightcurve peaks occurring at $\sim 87.5$ and $86$ days post-explosion, respectively \citep{2019apj...882l..15s}. Our RINGO3 observations straddle the lightcurve peak and the level of polarization observed ($\sim 0.7\%$) is slightly higher than seen for SN~1987A ($\sim 0.4\%$) at similar epochs \citep{1991apj...375..264j}.  Given the brightness of SN~2018hna, if we had been able to conduct earlier observations of this SN the RINGO3 observation would have been sufficiently sensitive to detect the presence of a similar peaks in the polarization that were seen for SN~1987A \citep{1991apj...375..264j}, if they occurred.

The observation of SN~2019nvm is one of the earliest polarimetric observations of a SN, potentially just beating the first observation by \citet{1987iauc.4319....1w} of SN 1987A.  \citet{2000apj...536..239l} reported early spectropolarimetry of the Type IIn SN 1998S 5 days after discovery (corresponding to $\sim 5 - 11$ days post-explosion), reporting levels of continuum polarization $\sim 2\%$.  Unlike the early emission-line features of SN~2013cu \citep{2014natur.509..471g}, the spectrum of SN1998S persisted for upto 14 days post-discovery, before cooling and developing the classical P Cygni profiles of a Type II SN \citep{2001mnras.325..907f, 2001apj...550.1030w}.  Although the spectral evolution of SN~1998S occurred over timescales of weeks, rather than days normally associated with ``flash" observations \citep{2014natur.509..471g, 2017natph..13..510y,2019apj...872..141s}, it does demonstrate that large levels of  polarization, due to presence of an aspherical CSM, could be observed at very early epochs.

This RINGO3 survey has crucially demonstrated the feasibility of employing polarimeters on robotic telescopes and, coupled with the appropriate feeder survey, the potential for ``flash polarimetry".  While it is expected that the CSM signature at early times would be erased by the expansion of the ejecta, in 7 cases we have been able to successfully observe targets within this 5 day window.  On 2 occasions, we have been able to trigger Liverpool Telescope observations within a day of discovery by ZTF.

The quality of the data presented in this paper has been limited by instrumental errors caused by the single-beam design of RINGO3 making the cancellation of systematic errors difficult.  RINGO3 was decommissioned on the Liverpool Telescope in January 2020 and replaced with a prototype of a new dual beam polarimeter (MOPTOP).  This uses a dual sCMOS imaging system to record the ordinary and extra-ordinary rays from a polarizing beam-splitter \citep{2016spie.9908e..4ij, 2018spie10702e..4qj}.  It therefore has higher throughput than the polaroid-based RINGO3 as well as allowing differential cancellation of polarization errors.   Commissioning observations with MOPTOP \citep{2020mnras.494.4676s} show uncorrected systematic errors reduced to $<0.2$ per-cent and the sensitivity increased by a factor $\sim 4 \times$ compared to RINGO3.  These combined improvements mean that MOPTOP has a polarization accuracy of $<0.3$\% for a source with $R=17$ in a 600 second exposure.  From our sample, 5 targets would be observable to this polarization precision, given that exposure time, at the earliest epochs; including SNe 2018hna and 2019ein.  MOPTOP will make tighter constraints or even detections of the polarization feasible for the types of targets we have observed so far with RINGO3.

The improved sensitivity of the MOPTOP instrument and the capability to observe in four wavelength bands, from the optical to the near-infrared, will potentially provide a better handle on the ISP.  As demonstrated for SN~2019ein, changes in the level of polarization are perceptible, and relative changes in intrinsic polarization can be directly measured.  A complete correction for the ISP could be derived, including inferring the wavelength dependence, if observations continued into later phases when a SN might be considered intrinsically unpolarized \citep[see e.g.][]{1991apj...375..264j} when, despite its faintness, it would still be accessible to MOPTOP.

Despite the constraints placed on the early-time explosion geometries with this RINGO3 programme, the limited sample size and only a limited number of detections of significant polarization means that ``flash polarimetry" is still {\it terra incognita}.  This survey has demonstrated it is possible to conduct these types of observations with very fast turnaround times and, with the advent of MOPTOP on the Liverpool Telescope, it will soon be possible to directly and systematically measure the polarization of SNe at the earliest times.

\section*{Acknowledgements}
The Liverpool Telescope is operated on the island of La Palma by Liverpool John Moores University at the Spanish Obervatorio del Roque de los Muchachos of the Instituto de Astrofisica de Canarias, with financial support from the UK Science and Technologies Facilities Council (STFC). Financial support for the development RINGO3 was also provided by the STFC Project Research and Development (PRD) scheme. 

Based on observations obtained with the Samuel Oschin Telescope 48-inch and the 60-inch Telescope at the Palomar Observatory as part of the Zwicky Transient Facility project. ZTF is supported by the National Science Foundation under Grant No. AST-1440341 and a collaboration including Caltech, IPAC, the Weizmann Institute for Science, the Oskar Klein Center at Stockholm University, the University of Maryland, the University of Washington, Deutsches Elektronen-Synchrotron and Humboldt University, Los Alamos National Laboratories, the TANGO Consortium of Taiwan, the University of Wisconsin at Milwaukee, and Lawrence Berkeley National Laboratories. Operations are conducted by COO, IPAC, and UW.

The research of JRM was supported through a Royal Society Research Fellowship.  The research of YY is supported through a Benoziyo Prize Postdoctoral Fellowship.  The research of AGY is supported by the EU via ERC grant No. 725161, the ISF GW excellence center, an IMOS space infrastructure grant and BSF/Transformative and GIF grants, as well as The Benoziyo Endowment Fund for the Advancement of Science, the Deloro Institute for Advanced Research in Space and Optics, The Veronika A. Rabl Physics Discretionary Fund, Paul and Tina Gardner, Yeda-Sela and the WIS-CIT joint research grant;  AGY is the recipient of the Helen and Martin Kimmel Award for Innovative Investigation.  The work of X. Wang is supported by National Natural Science Foundation of China (NSFC grants 11325313, 11633002, and 11761141001), and the National Program on Key Research and Development Project (grant no. 2016YFA0400803).  The authors thank S. Van Dyk for useful comments on the manuscript.

\section*{Data Availability}
The observational data presented here is available in the public archive of the Liverpool Telescope (\url{https://telescope.livjm.ac.uk/cgi-bin/lt_search}) 
%%%%%%%%%%%%%%%%%%%%%%%%%%%%%%%%%%%%%%%%%%%%%%%%%%

%%%%%%%%%%%%%%%%%%%% REFERENCES %%%%%%%%%%%%%%%%%%

% The best way to enter references is to use BibTeX:

\bibliographystyle{mnras}

%%%%%%%%%%%%%%%%%%%%%%%%%%%%%%%%%%%%%%%%%%%%%%%%%%

%%%%%%%%%%%%%%%%% APPENDICES %%%%%%%%%%%%%%%%%%%%%

%\appendix

% Don't change these lines
\bsp	% typesetting comment
\label{lastpage}
\end{document}